\documentclass[12pt]{article}

\newcommand{\eq}{\begin{equation}}
\newcommand{\eqx}{\end{equation}}
\newcommand{\eqn}{\begin{eqnarray}}
\newcommand{\eqnx}{\end{eqnarray}}
\newcommand{\f}[2]{\frac{#1}{#2}}

\newcommand{\tr}{\mbox{\rm tr}\,}
\newcommand{\dl}{\delta}

\newcommand{\dery}{\f{\partial}{\partial Y}}
\newcommand{\der}{\partial}
\newcommand{\NN}{{\cal N}}
\newcommand{\CS}{{\cal S}}
\newcommand{\HH}{{\cal H}}
\newcommand{\DD}{{\cal D}}
\newcommand{\cor}[1]{\left\langle{#1}\right\rangle}

\title{QCD saturation in the dipole sector with correlations}

\author{Romuald A. Janik\footnote{e-mail: {\tt
      ufrjanik@if.uj.edu.pl}}\\ \\
Institute of Physics\\ 
Jagellonian University\\ 
ul. Reymonta 4\\
30-059 Krakow\\ 
Poland}

\begin{document}

\maketitle

\begin{abstract}
In this paper we study the description of saturation in   
Balitsky, Jalilian-Marian, Iancu, McLerran, Weigert,
Leonidov and Kovner (B-JIMWLK) picture when restricted to observables
made up only from dipole operators.
We derive a functional form of the evolution equation for the dipole
probability distribution
and find a one-parameter family of exact solutions to the dipole evolution
equations. 
\end{abstract}

\section{Introduction}

One of the outstanding issues in perturbative QCD is the understanding
of parton saturation. Standard evolution equations like BFKL
\cite{BFKL} lead to
rising cross sections (or structure functions) which violate the
unitarity bound. These are modified, when one takes into account
recombination effects, into an infinite set of coupled nonlinear
equations for correlation functions of Wilson line operators
\cite{BALITSKY}. This set of equations is expected to be equivalent to
the Jalilian-Marian, Iancu, McLerran, Weigert, Leonidov and Kovner
(JIMWLK) functional equations \cite{JIMWLK}, which can be recast in
the form of Fokker-Planck equations for the (functional) probability
distribution of Wilson lines $Z_Y[\{ U_x \}]$. The probability
distribution $Z_Y[\{ U_x \}]$ encodes expectation values of
observables made up of the longitudinal Wilson lines $U_x$, where $x$
is a coordinate in the transverse space: 
\eq
\cor{f[\{ U_x \}]} \equiv \int \DD U_x \,f[\{ U_x \}]\, Z_Y[\{ U_x \}]
\eqx
The full knowledge of $Z_Y[\{ U_x \}]$ is of course nonperturbative,
but the {\em evolution} in rapidity is governed by the JIMWLK
functional equation:
\eq
\label{e.jimwlk}
\dery Z_Y[\{ U_x \}] = \alpha_s \f{1}{2}\int d^2x\, d^2y\, \left( i
\nabla^a_{U_x} \,\hat{\chi}^{ab}_{xy}\, i \nabla^b_{U_y} \right) \,
Z_Y[\{ U_x \}] 
\eqx
where the derivatives $\nabla^a_{U_x}$ are covariant
functional derivatives w.r.t. the unitary matrices while the kernel
$\hat{\chi}^{ab}_{xy}$ is given by the following expression:
\eq
\hat{\chi}^{ab}_{xy} =\f{1}{\pi^2} \int d^2z \left(
\f{(x-z)_i}{(x-z)^2} [1-U_x^{-1} U_z]^{ac} \right)
\left(\f{(z-y)_i}{(z-y)^2} [1-U_z^{-1} U_y]^{cb} \right)
\eqx 
See \cite{WEIGERT} and \cite{BI} for further details.
Despite the apparent compactness of the above equation which
summarizes the whole nonlinear infinite set of evolution equations for
correlation functions it has proved itself to be
difficult to analyze. Apart from some general properties
not much is explicitly known about the details of
saturation/unitarization in this picture.

It is interesting, therefore,  to study simplified versions. One
noteworthy example is the Balitsky-Kovchegov (BK) equation
\cite{BALITSKY,Kovchegov} 
obtained when correlation functions of traces of products Wilson lines
factorize into products of expectation values. This leads to a
nonlinear equation for the dipole density
\eq
\dery \cor{\CS_{01}} = \f{g^2}{8 \pi^3} \int d^2x_2 \, 
\f{x_{01}^2}{x_{02}^2 x_{21}^2} 
\left[\cor{\CS_{02}}\cor{\CS_{21}} - N_c \cor{\CS_{01}} \right]
\eqx
where the $\CS_{ij}$ are (the S-matrix versions of) the dipole
operators
\eq
\CS_{ij}=\tr U_{x_i} U^\dagger_{x_j}\ ,
\eqx
The Balitsky-Kovchegov equation was studied both numerically and analytically
\cite{BKlit} and boasts
universal properties of geometrical scaling \cite{GEOM} (appearing
here as universal traveling wave solutions \cite{ROBI}).

In order to better understand the full hierarchy of JIMWLK equations,
it is of particular interest to find some intermediate approximations
which are not so drastic as BK, yet remain simpler than the full
JIMWLK hierarchy. One approach might be to restrict oneself just to
observables made up from dipole operators, as these have a very clear
intuitive physical meaning, yet to refrain from making any other
simplifying assumptions. 
The generalization with respect to BK means that one has to allow for
nontrivial correlations between dipoles i.e. allowing for
$\cor{\CS_{01}\CS_{23}} \neq \cor{\CS_{01}}\cor{\CS_{23}}$ etc. In
\cite{US} we studied the first step beyond BK with nontrivial 2-dipole
correlations. Other approaches going beyond BK include
\cite{LEVIN,IANCU,SHOSHI}.

The aim of this paper is to derive an analogue of (\ref{e.jimwlk})
restricted to probability distributions made up of dipole operators
only, and to find a one-parameter class of {\em exact} solutions of
these dipole evolution equations with some specific multi-dipole
correlations.

The comparison  of the dipole evolution hamiltonian to the full JIMWLK
one may be a first step in understanding the role of higher multipole
operators in perturbative unitarization/saturation processes.

The plan of this paper is as follows. In section 2 we derive the
evolution hamiltonian for the JIMWLK hierarchy restricted to dipole
operators, then in section 3 we discuss some of its properties. 
In section 4 we construct a one-parameter solution of the dipole
evolution equations and we close the paper with a summary and
discussion. 

\section{The JIMWLK hierarchy restricted to dipoles}

In order to calculate the evolution of an observable of the form
$\cor{\CS_{01}\CS_{23}\ldots \CS_{n-1\,n}}$ it is most convenient to
use Balitsky's approach where the evolution with rapidity is obtained
using contractions between pairs of the unitary matrices $U_x$'s
appearing in the dipole operators and virtual corrections for each
$U_x$ according to eqs. (119), (120) in \cite{BALITSKY}. One can
verify that contractions between $U_x$'s belonging to different
dipoles always give rise to multipoint traces, so if one wants to keep
only terms involving dipoles one performs contractions only within
each dipole and therefore each dipole operator evolves independently
through the {\em operator} form of the BK kernel:
\eq
\dery f[\CS_{xy}]=  \f{g^2}{8\pi^3} \int d^2x\, d^2y\, d^2z\,
\left(\f{\dl}{\dl S_{xy}} f[\CS_{xy}] \right)  K_{xzy} (\CS_{xz}
\CS_{zy} - N_c \CS_{xy})
\eqx
Introducing the dipole probability distribution $Z_Y[\{\CS_{xy}\}]$ which
encodes expectation values through
\eq
\label{e.defz}
\cor{f[\{ \CS_{xy} \}]} \equiv \int \DD \CS_{xy} \,f[\{ \CS_{xy} \}]\,
Z_Y[\{ \CS_{xy} \}] 
\eqx
we can rewrite the previous equation as
\eqn
\label{e.a}
\dery \cor{f[\CS_{xy}]} &=& \f{g^2}{8\pi^3} \int {\cal D} \CS_{xy}\int
d^2x d^2y d^2z \left(\f{\dl}{\dl S_{xy}} f[\CS_{xy}] \right) \cdot
\nonumber\\
&& \cdot K_{xzy} (\CS_{xz} \CS_{zy} - N_c \CS_{xy}] Z_Y[\{\CS_{xy}\}]
\eqnx
We will now recast the right hand side as
\eq
\int {\cal D} \CS_{xy} \,f[\CS_{xy}] \,\HH Z_Y[\{\CS_{xy}\}]
\eqx
Then the evolution equation for the dipole probability distribution
will have the form
\eq
\dery Z_Y[\{\CS_{xy}\}] = \HH Z_Y[\{\CS_{xy}\}]
\eqx
If we perform a functional integration by parts in (\ref{e.a}) and assume the
vanishing of boundary terms we obtain an explicit and quite simple
form of the evolution hamiltonian $\HH$:
\eq
\label{e.dipoleeq}
\dery Z_Y[\{\CS_{xy}\}] =  \f{-g^2}{8\pi^3} \int d^2x d^2y d^2z
\f{\dl}{\dl S_{xy}}  \left\{ K_{xzy} (\CS_{xz} \CS_{zy} - N_c
\CS_{xy}) \right\} Z_Y[\{\CS_{xy}\}] 
\eqx

\section{Some remarks on the dipole hamiltonian}

As an example of how the functional evolution equation
(\ref{e.dipoleeq}) works in practice let us quickly rederive the
equation for 2-dipole correlations obtained in \cite{US}. Let us start
from the identity
\eq
\dery \cor{\CS_{01} \CS_{23}}= \int \DD \CS_{xy}\, \CS_{01} \CS_{23}
\, \dery Z_Y[\{\CS_{xy}\}] 
\eqx
Now we use (\ref{e.dipoleeq}) and perform integration by parts to get
\eq
\f{g^2}{8\pi^3} \int \DD \CS_{xy} \int d^2x d^2y d^2z \f{\dl}{\dl
S_{xy}} \left( \CS_{01} \CS_{23} \right) \,\left\{ K_{xzy} (\CS_{xz}
\CS_{zy} - N_c \CS_{xy}) \right\} Z_Y[\{\CS_{xy}\}] 
\eqx
After performing the functional differentiation and using the definition of
$Z_Y[\{\CS_{xy}\}]$ (\ref{e.defz}) we get the equation obtained in
\cite{US}: 
\eqn
\dery \cor{\CS_{02}\CS_{2'1}} &=&\f{g^2}{8\pi^3} \int d^2 x_3 \left[
  \cor{\CS_{03} \CS_{32} \CS_{2'1}} -N_c \cor{\CS_{02}\CS_{2'1}}
  \right] K_{032}+ \nonumber \\  
&&+\left[
  \cor{\CS_{2'3} \CS_{31} \CS_{02}} -N_c \cor{\CS_{02}\CS_{2'1}}
  \right] K_{132'} \ .
\eqnx

Let us now contrast some properties of the dipole evolution
equation (\ref{e.dipoleeq}) with the full JIMWLK one
(\ref{e.jimwlk}). The latter one is second order in functional
derivatives and, as shown by Weigert, the hamiltonian is positive
definite, with a {\em unique} attractive fixed point ($Z_Y[\{U_x\}]=1$). 
On the other hand the dipole equation is first order and as
such does not seem to have such positivity and uniqueness
properties. Even more so, from the form of (\ref{e.dipoleeq}) one
would expect that an infinite set of zero-modes would exist.
This would lead to many probability distributions determined by
\eq
\HH Z_Y[\{\CS_{xy}\}]=0
\eqx
which are invariant under evolution.

We cannot {\em a-priori} rule this out but let us show on a simple toy
model 
how one can nevertheless escape this conclusion.
Assume the following simple evolution equation for a dipole operator
\eq
\dery s=s-s^2
\eqx  
which exhibits saturation behaviour quite similar to BK.
This leads to the hamiltonian for the probability distribution $Z_Y[s]$:
\eq
\dery Z_Y[s]=\der_s \left[ (s^2-s)Z_Y[s] \right]
\eqx
One can now find explicitly the dangerous stationary solutions:
\eq
Z_Y[s]=\f{c}{s(1-s)}
\eqx
However these solutions are nonphysical as they are nonnormalizable
due to the singularities at $s=0,1$, and therefore cannot represent any
probability distribution.
It would be very interesting (but rather difficult) to carry out a similar
analysis for the real dipole equation (\ref{e.dipoleeq}).

\section{A family of factorized solutions to the dipole hierarchy of
  evolution equations}

It is interesting to look for {\em exact} solutions of the full set of
dipole evolution equations. It turns out to be easier to find a
solution of the type considered here using directly a set of evolution
equations for the moments of dipole operators rather than finding
first the probability distribution $Z_Y[\{\CS_{xy}\}]$.

The JIMWLK-Balitsky hierarchy of evolution equations, when restricted
to dipole operators has the following form:
\eq
\label{e.hs}
\dery \CS_1 \CS_2 \ldots \CS_n= \sum_{i=1}^n \CS_1 \ldots \CS_{i-1}
(\HH \CS_i) \CS_{i+1}\ldots \CS_n
\eqx
In terms of dipole density operators $\NN_i$ related to the $\CS_i$
through
\eq
\CS_i=N_c(1-\NN_i)
\eqx 
the hamiltonian acts as
\eq
\label{e.hamilt}
\HH \NN_i \equiv \HH \NN_{ii'} = \f{g^2 N_c}{8\pi^3} \int d^2x K_{ixi'}\left[
\NN_{ix}+ \NN_{xi'} -  \NN_{ii'} -\NN_{ix}\NN_{xi'}  \right]
\eqx

Let us look for solutions to this set of coupled equations in a
factorized form:
\eq
\label{e.fact}
\cor{\NN_1 \NN_2\ldots \NN_n} = c_n \cor{\NN_1}\cor{\NN_2}\ldots
\cor{\NN_n} \equiv c_n N_1 N_2 \ldots N_n
\eqx
with $c_n$ some coefficients ($c_1\equiv 1$). Note that $\cor{\ldots}$
is the ordinary
expectation value and not just the connected part. If all $c_n$ would
be equal to one, this would imply no nontrivial correlations and would
be equivalent to the standard BK equation.

Let us take the expectation value in the evolution equation (\ref{e.hs}):
\eqn
\label{e.neq}
\dery \cor{(1-\NN_1)\ldots (1-\NN_n)}\!\!\!\! &=&\!\!\!\!
-\sum_{i=1}^n \biggl\langle (1-\NN_1) \ldots (1-\NN_{i-1}) (\HH \NN_i)
(1-\NN_{i+1}) \nonumber\\ 
&&\ldots (1-\NN_n)\biggr\rangle
\eqnx
We may now use first the assumption of factorization (\ref{e.fact}) in the
{\em left hand side} of (\ref{e.neq}). Thus the $Y$-derivative acts on
products of single dipole densities $N_i=\cor{\NN_i}$. Then for each
$N_i$ we use the evolution operator
\eq
\dery N_1 =\f{g^2 N_c}{8\pi^3} \int d^2x K_{ixi'}\left[N_{ix}+
N_{xi'} -  N_{ii'} -c_2 N_{ix} N_{xi'}  \right]
\eqx
Note the appearance of $c_2$ w.r.t. (\ref{e.hamilt}) as we are taking
expectation values.
We can now insert the {\em operator form} of (\ref{e.hamilt}) into the
{\em right hand side} of (\ref{e.neq}) and then perform the averaging using the
factorization ansatz (\ref{e.fact}). Comparing terms we see that all
terms match if the following holds:
\eq
c_{i+1}=c_2 c_i
\eqx
which has a simple solution\footnote{After these results were obtained
M. Lublinsky informed me that the same factorized solution appeared in a
framework of generalized BK in \cite{LEVIN}.} 
\eq
\label{e.sol}
c_n=c_2^{n-1}
\eqx
Therefore all multidipole correlation functions reduce to one-dipole
densities but with an {\em enhancement} factor:
\eq
\label{e.gensol}
\cor{\NN_1 \NN_2\ldots \NN_n} = c_2^{n-1} N_1 N_2 \ldots N_n
\eqx
And all the one dipole densities $N_i$ satisfy a {\em modified} BK
equation
\eq
\dery N_{ii'} =\f{g^2 N_c}{8\pi^3} \int d^2x K_{ixi'}\left[N_{ix}+
N_{xi'} -  N_{ii'} -c_2 N_{ix} N_{xi'}  \right]
\eqx
with $c_2$ being an {\em a-priori} arbitrary parameter. However in order to have a solution which saturates at $N_i \leq 1 $, $c_2$ has to be greater or equal to 1. Indeed let us note that (\ref{e.gensol}) with $N_i=1/c_2$ is a fixed point of the dipole evolution equations. It would be interesting to recast this solution in the language of probability distributions $Z_Y[\{\CS_{xy}\}]$ and analyze its normalizability properties. 

We note that the same modified equation was proposed in \cite{LEVIN}
to acccount for effects of correlations in
nuclei on saturation. It is interesting to note that the same equation
arises from an exact solution of the dipole evolution hierarchy.

Let us note how does the factorized solution of \cite{US} fit into the above
general framework. That solution was obtained assuming that the dipole
evolution equations close at the level of two equations for
$\cor{\NN_1}$ and $\cor{\NN_1\NN_2}$ i.e. that there are
no nontrivial 3-dipole correlations. In order to recover that solution
in our framework we have to analyze the {\em connected} correlation
functions 
\eq
\cor{\NN_1 \NN_2\ldots \NN_n}_c = d_n N_1 N_2 \ldots N_n
\eqx
The coefficients $d_n$ are related to the $c_n$'s through
\eqn
c_2 &=& c_1^2 +d_2 \\
c_3 &=& c_1^3+3c_1 d_2 +d_3 \\
c_4 &=& c_1^4 +6d_2 c_1^2 +3 d_2^2 +4 c_1 d_3 +d_4
\eqnx
etc. The requirement that connected 3-dipole correlations vanish leads
to the equation $d_3=c_2^2-1-3(c_2-1)=0$ which fixes $c_2$ to be
either 1 (no correlations at all) or $c_2=2$ (the solution found in
\cite{US}). We note that from the general formula (\ref{e.sol}) that
solution (for $\cor{\NN_1}$ and $\cor{\NN_1\NN_2}$) can be extended to
a solution of the full dipole hierarchy but with some nonvanishing
higher-dipole correlations (like $d_4\neq 0$).  

\section{Summary}

In this paper we studied the JIMWLK evolution equations in the
approxmation of keeping only dipole-like color contractions and
neglecting all higher multipole operators. In this approximation we
have derived an equation for the evolution of the dipole probability
distribution. No further assumptions like neglecting correlations were
made. The hamiltonian turned out to be first order in functional
derivatives in contrast to the full JIMWLK evolution. It would be
interesting to study further properties of this hamiltonian.

In the second part of the paper we derived a one-parameter family of
exact solutions to the dipole evolution hierarchy. They are
characterized by an enhancement (or depletion) factor in multidipole
correlation functions and the single dipole densities are governed by
a {\em modified} BK equation.

It seems that it would be very interesting to study further the
properties of saturation when restricted to the dipole sector since
this is a much simpler system than the full JIMWLK one. In particular
one could explore the probabilisitic interpretation of the dipole
evolution equations (\ref{e.dipoleeq}), the structure of (attractive)
fixed points and the question whether multipole operators present in
the full JIMWLK framework lead to qualitative or just quantitative
changes in the physics of saturation. 

\medskip

\noindent{\bf Acknowledgements.} I would like to thank Edmond Iancu,
Michael Lublinsky and Robi Peschanski for discussions and
comments. This work was supported in part by KBN grants 2P03B09622
(2002-2004), 2P03B08225 (2003-2006) and 1P03B02427 (2004-2007).


\begin{thebibliography}{10}

\bibitem{BFKL}
L.~N. Lipatov,
\newblock Sov. J. Nucl. Phys. {\bf 23}, 338 (1976);
\newblock E.~A. Kuraev, L.~N. Lipatov, and V.~S. Fadin,
\newblock Sov. Phys. JETP {\bf 45}, 199 (1977);
\newblock I.~I. Balitsky and L.~N. Lipatov,
\newblock Sov. J. Nucl. Phys. {\bf 28}, 822 (1978).

\bibitem{BALITSKY}
I.~Balitsky,
\newblock  Nucl. Phys.  {\bf B463} (1996) 99.

\bibitem{JIMWLK}J.Jalilian-Marian,A. Kovner, A. Leonidov and H. Weigert,
\newblock  Nucl. Phys. {\bf  B504} (1997) 415; Phys. Rev. {\bf D59}  (1999) 
014014.	
H. Weigert, 
\newblock Nucl. Phys. {\bf A703} (2002) 823.	E.Iancu, A. Leonidov and L. 
McLerran, 
\newblock Nucl.Phys. {\bf A692}  (2001) 583; Phys.Lett. {\bf B510}  (2001) 133.

\bibitem{WEIGERT}
H.~Weigert,
Nucl.\ Phys.\ A {\bf 703} (2002) 823
[arXiv:hep-ph/0004044].

\bibitem{BI}
J.~P.~Blaizot, E.~Iancu and H.~Weigert,
Nucl.\ Phys.\ A {\bf 713}, 441 (2003)
[arXiv:hep-ph/0206279].

\bibitem{Kovchegov}
Y.~V. Kovchegov,
\newblock Phys. Rev. {\bf D60}, 034008 (1999); Phys. Rev. {\bf D61}, 074018 
(2000).

\bibitem{BKlit} See e.g. 
K.~Golec-Biernat and A.~M.~Stasto,
Nucl.\ Phys.\ B {\bf 668} (2003) 345
[arXiv:hep-ph/0306279].
A.~H.~Mueller and D.~N.~Triantafyllopoulos,
Nucl.\ Phys.\ B {\bf 640} (2002) 331
[arXiv:hep-ph/0205167].
E.~Iancu, K.~Itakura and L.~McLerran,
Nucl.\ Phys.\ A {\bf 708}, 327 (2002)
[arXiv:hep-ph/0203137].
J.~Bartels, E.~Gotsman, E.~Levin, M.~Lublinsky and U.~Maor,
Phys.\ Rev.\ D {\bf 68} (2003) 054008
[arXiv:hep-ph/0304166].


\bibitem{GEOM}
A.~M. Sta\'sto, K.~Golec-Biernat, and J.~Kwiecinski,
\newblock Phys. Rev. Lett. {\bf 86}, 596 (2001), hep-ph/0007192.

\bibitem{ROBI}
S.~Munier and R.~Peschanski,
Phys.\ Rev.\ Lett.\  {\bf 91}, 232001 (2003); 
Phys.\ Rev.\ D {\bf D69}, 034008 (2004); hep-ph/0401215.

\bibitem{US}
R.~A.~Janik and R.~Peschanski,
arXiv:hep-ph/0407007.


\bibitem{LEVIN}
E.~Levin and M.~Lublinsky,
Nucl.\ Phys.\ A {\bf 730} (2004) 191
[arXiv:hep-ph/0308279].

\bibitem{IANCU}
E.~Iancu and A.~H.~Mueller,
Nucl.\ Phys.\ A {\bf 730}, 494 (2004)
[arXiv:hep-ph/0309276].

\bibitem{SHOSHI}
A.~H.~Mueller and A.~I.~Shoshi,
Nucl.\ Phys.\ B {\bf 692} (2004) 175
[arXiv:hep-ph/0402193].

\end{thebibliography}
\end{document}